\documentclass
[prd,twocolumn,showpacs,preprintnumbers]{revtex4}
\usepackage{amssymb}
\usepackage{graphicx}
\usepackage{dcolumn}
\usepackage{bm}
\usepackage{amsfonts}

\newcommand{\n}{{\cal A}}
\newcommand{\A}{A}
\newcommand{\B}{B}
\newcommand{\C}{C}
\newcommand{\D}{D}

\newcommand{\td}{{\tilde d}}

\newcommand{\e}{{\rm e}}
\newcommand{\de}{ \Delta}

\newcommand{\s}{{\sigma}}

\newcommand{\lb}{\label}
\newcommand{\be}{\begin{equation}}
\newcommand{\ee}{\end{equation}}
\newcommand{\bea}{\begin{eqnarray}}
\newcommand{\eea}{\end{eqnarray}}
\newcommand{\bw}{\begin{widetext}}
\newcommand{\ew}{\end{widetext}}
\begin{document}

\preprint{DTP-MSU/04-06} \preprint{ DF/IST-03.2004}
\preprint{LAPTH-1030/04}

\preprint{hep-th/0403112}

\title{Supergravity $p$-branes revisited: extra parameters,
uniqueness, and topological censorship}

\author{Dmitri V. Gal'tsov}
 \email{galtsov@grg.phys.msu.su}
 \affiliation{Department of Theoretical Physics, \\Moscow State
University,
119899, Moscow, Russia}

\author{Jos\'e P. S. Lemos}
 \email{lemos@kelvin.ist.utl.pt}
 \affiliation{CENTRA, Departamento de F\'{\i}sica,
              Instituto Superior T\'ecnico,\\
   Av. Rovisco Pais 1, 1096 Lisboa, Portugal}

\author
{G\'erard Cl\'ement} \email{gclement@lapp.in2p3.fr}
\affiliation{Laboratoire de  Physique Th\'eorique LAPTH (CNRS),\\
B.P.110, F-74941 Annecy-le-Vieux cedex, France }

\begin{abstract}
We perform a complete integration of the
Einstein-dilaton-antisymmetric form action describing black
$p$-branes in arbitrary dimensions assuming the transverse space
to be homogeneous and possessing spherical, toroidal or hyperbolic
topology. The generic solution contains eight parameters
satisfying one constraint. Asymptotically flat solutions form a
five-parametric subspace, while conditions of regularity of the
non-degenerate event horizon further restrict this number to
three, which can be related to the mass and the charge densities
and the asymptotic value of the dilaton. In the case of a
degenerate horizon, this number is reduced by one. Our derivation
constitutes a constructive proof of the uniqueness theorem for
$p$-branes with the homogeneous transverse space. No
asymptotically flat solutions with toroidal or hyperbolic
transverse space within the considered class are shown to exist,
which result can be viewed as a demonstration of the topological
censorship for $p$-branes. From our considerations it follows, in
particular, that some previously discussed $p$-brane-like
solutions with extra parameters do not satisfy the standard
conditions of asymptotic flatness and absence of naked
singularities. We also explore the same system in presence of a
cosmological constant, and derive a  complete analytic solution
for higher-dimensional charged topological black holes, thus
proving their uniqueness.
\end{abstract}


\maketitle

\section{Introduction}

Classical solutions of the supergravity equations, describing
$p$-branes charged with respect to  a $p+1$ antisymmetric form
field, were extensively studied during the past decade
\cite{HoSt91,Gu92,DuLu94,LuPoSeSt95,LuPo95,LuPoXu95,DuKhLu95,DuLuPo96,LuPoSeSt96,St98,
GaRy98,IvMe01}. In the case of a single brane (which will be the
only one discussed here), the standard (black) brane solution
depends on two parameters, the mass and the charge (densities) of
the brane \cite{HoSt91,DuLu94}, these solutions are asymptotically
flat and possess a regular event horizon. Black $p$-branes have
the $ISO(p)\times R$ symmetry of the world volume (with $R$
corresponding to the time direction), which is enhanced to the
full Poincar\'e symmetry $ISO(p,1)$ in the extremal (BPS) case. In
the simplest case the  space transverse to the brane is taken to
be spherically symmetric, the generalizations to the products of
the lower-dimensional sphere by the flat space are also known.

Both the BPS and the black branes were first obtained by solving
the corresponding field equations under special ans\"atze for the
metric apart from the above symmetries \cite{DuLu94,DuKhLu95}, so
the degree of generality of these solutions was not clear a
priori. Alternatively, they may be  generated from a suitably
smeared Schwarzschild solution by Harrison-type transformations
\cite{GaRy98}.  The question of the uniqueness of the standard
$p$-brane solutions was raised recently \cite{Gi02}, but the
explicit proof was given only in the case  $p=0$, that is for
multidimensional black holes, and only for a non-degenerate event
horizon. Meanwhile, some more general solutions to the same system
depending on higher number of parameters were also suggested
\cite{LuPoXu95,IvMe01} for $ISO(p,1)$ symmetric branes. More
recently a complete integration of the
Einstein-dilaton-antisymmetric form system for a single brane was
performed \cite{ZhZh99} and a family of $ISO(p)\times R$ solutions
was presented, containing four free parameters. An interpretation
of one of the extra parameters was attempted in \cite{BrMaOz01}
(see also \cite{BrSt01}): the $ISO(p,1)$ subfamily of the
solutions of \cite{ZhZh99} was treated as describing a
brane-antibrane system in the sense of Sen \cite{Se99,LeRu99}, the
corresponding extra degree of freedom  being associated with the
tachyon. The solutions of \cite{ZhZh99} were also invoked in some
recent attempts to find a supergravity description for stable
non-BPS branes in string theory \cite{BeVe00,BeLe00,Lo01,Ba01}.
Without entering into a discussion of the consistency of these
suggestions, we would like here to stress that the detailed
 structure of singularities of the $p$-brane-like solutions with extra
parameters was not investigated so far.  Other generalizations of
the $ISO(p,1)$ solution were given in \cite{ChGaGu02}. Besides
containing additional parameters, the solutions presented in this
paper also describe a more general structure of the transverse
space, namely, $SO(k)\times R^{q},\; q=D-p-k-2$ (cylinder),
$R^{q+k+1},\;$ as well as the case of the hyperbolic geometry
$SO(k-1,1)\times R^{q},\;$. The latter two cases were previously
explored  for $p=0$ (topological black holes) in the presence of a
negative cosmological constant, causing the space-time to possess
an asymptotic AdS structure
\cite{Le94,Le95,Le96,Ca96,Va97,Br97,Ma97,Kl98,Bi98,Ca99,Pe00,Le00}.
It can be assumed that the flat and the hyperbolic spaces are
compactified by factoring the space by a discrete subgroup of the
isometry group \cite{Br97} and describe black holes with toroidal
and higher genus surfaces of the event horizon (for a recent
review see \cite{Le00}). Whether such solutions could be extended
to $p$-branes remains unclear, since no analytic solutions for
$p$-branes with cosmological constant were found so far
\cite{Du99}. Neither are we aware of any topological censorship
arguments for $p$-branes of the  type known for black holes
\cite{Ga99,CaGa01}.

The aim of this paper is to clarify the nature of generic
solutions to  Einstein-dilaton-antisymmetric form equations
describing $p$-branes with spherical, toroidal and hyperbolic
geometry of the transverse space. We construct generic solutions
by reducing the system to two Liouville equations, and then
carefully investigate restrictions on the free parameters
following from the requirements of the asymptotic flatness and the
absence of naked singularities. We find that in both cases of
non-degenerate, or degenerate event horizons, the generic solution
is reduced effectively to the standard black brane or the BPS
solution, fully characterized by the tension, the charge density
and the asymptotic value of the dilaton (this parameter was often
ignored in the previous literature). No extra parameters can be
added without violating the regularity conditions. This proves the
uniqueness theorem for $p$-branes for both non-degenerate and
degenerate event horizons, in the case of $ISO(p)\times R$ and
$ISO(p,1)$ symmetries of the world-volume and a spherical
transverse space. We also prove the absence of asymptotically flat
$p$-brane solutions with toroidal and hyperbolic transverse
geometries, thus generalizing the topological censorship
conjecture for $p$-branes. We also explore the same system in the
presence of a cosmological constant, in the hope that topological
$p$-brane solutions will arise. But in this case the system is not
fully integrable, and we were only able to give a simple
constructive derivation of the topological $0$-brane with AdS
asymptotics.

\section{General  setting}
We consider the  action  containing a graviton, a $q$-form field
strength F$_{[q]}$, and a dilaton scalar $\phi$ coupled to the
form field with  coupling constant $a$. This is a general
framework which encompasses the bosonic sector of various bosonic
models, coming from a truncation of the low energy limit of
M-theory and string theories, for a certain choice of the
dimension $D$, the rank $k$ of the form field, and the dilaton
coupling $a$. In the Einstein frame, the action is given by
\begin{equation}\label{action}
S = \int d^D x \sqrt{-g} \left( R - \frac12 \partial_\mu \phi
\partial^\mu \phi - \frac1{2\, k!} \, {\rm e}^{a\phi} \, F_{[k]}^2
\right).
\end{equation}
The corresponding field equations are invariant under the
following discrete S-duality:
\begin{equation} \label{duality}
g_{\mu\nu} \to g_{\mu\nu}, \qquad F \to {\rm e}^{-a\phi} \ast F,
\qquad \phi \to -\phi,
\end{equation}
where $\ast$ denotes the $d$-dimensional Hodge dual. This may be
used to construct electric versions of magnetic $S$-branes and
vice versa, so here we will consider only magnetic solutions. The
equations of motion, derived from the variation of the action with
respect to the individual fields, are \bw
\begin{eqnarray}
R_{\mu\nu} - \frac12 \partial_\mu \phi \partial_\nu \phi -
\frac{{\rm e}^{a\phi}}{2(k-1)!} \left[
F_{\mu\alpha_2\cdots\alpha_k} F_\nu{}^{\alpha_2\cdots\alpha_k}-
\frac{k-1}{k(D-2)} F_{[k]}^2 \, g_{\mu\nu} \right] &=& 0,
\label{Ein} \\
\partial_\mu \left( \sqrt{-g} \, {\rm e}^{a\phi} \,
F^{\mu\nu_2\cdots\nu_k} \right) &=& 0, \label{form} \\
\frac1{\sqrt{-g}}\, \partial_\mu \left( \sqrt{-g} \partial^\mu
\phi \right) - \frac{a}{2\, k!} {\rm e}^{a\phi} F_{[k]}^2 &=& 0.
\label{dil}
\end{eqnarray}
\ew We study $p$-branes with a world volume given by a $p+1$
dimensional space with isometries $ISO(p)\times R$ and with a
transverse space being the $\td+1$ dimensional space
$\Sigma_{k,\sigma}$. We will use the standard notation for
dimensions $d=p+1,\;\td=k-1=D-d-2$.  The space-time interval
\begin{equation}\label{metric}
ds^2 = - {\rm e}^{2\B} dt^2 + {\rm e}^{2\D} (dx_1^2 + \cdots +
dx_p^2) + {\rm e}^{2\C} \, d\Sigma_{k,\sigma}^2 + {\rm e}^{2\A}
dr^2,
\end{equation}
is parameterized by four functions $\A(r),\,\B(r),\, \C(r)$ and
$\D(r)$ depending only on $r$. The transverse space
$\Sigma_{k,\sigma}$ for $\sigma=0,+1,-1$ is the $k$-dimensional
flat space, the sphere and the hyperbolic space respectively:
\begin{equation}\lb{trans}
d\Sigma_{k \sigma}^2 = \bar g_{ab} dy^a dy^b = \left\{
 \begin{array}{ll}
d \psi^2 + \sinh^2\psi \, d\Omega_{\td}^2, \;\;&
\sigma=-1,\nonumber\\
d \psi^2 + \psi^2 \, d\Omega_{\td}^2, \;\; & \sigma=0,\\
d \psi^2 + \sin^2\psi \, d\Omega_{\td}^2,\;\; &
\sigma=1,\nonumber\\
 \end{array} \right.
\label{gmetric}\end{equation} The Ricci tensor for the transverse
space reads
\begin{equation}
\bar R_{ab} = \sigma \td \bar g_{ab}.
\end{equation}
The metrics (\ref{trans})  have $SO(\td,1),\, ISO(k)$ and $SO(k)$
isometries respectively.  In the flat and hyperbolic case one can
assume suitable compactifications by factoring over an appropriate
discrete subgroup of the isometry group, e.g. for $\sigma=0$ one
can choose a torus, for $\sigma=-1$ -- some compact hyperbolic
space \cite{KMST}.

With this ansatz, the equation for the form field (\ref{form}),
can be easily be solved,
\begin{equation}\label{SolF}
F_{[k]} = b  \,\, \mbox{vol}(\Sigma_{k,\sigma}),
\end{equation}
where $b$ is the field strength parameter, and
$\mbox{vol}(\Sigma_{k,\sigma})$ denotes the volume form of the
space $\Sigma_{k,\sigma}$.

  The Ricci tensor for the metric
(\ref{metric}) has the non-vanishing components
\begin{eqnarray}
&&R_{tt}={\rm e}^{2\B-2\A} \left(   \B'' + \B'F'    \right),
\label{Rtt}\\
&&R_{xx}=-{\rm e}^{2\D-2\A}\left( \D'' + \D'F'\right) ,\\
&&R_{rr}=-F''- \A'' - \B'(\B'-\A')-\nonumber\\&&-(\td +1)\C'(
\C'-\A')-(d-1)D'(\D'-\A'), \\
&&R_{ab}=\left(-{\rm e}^{2\C-2\A}\left(\C''+\C'F'\right)+\sigma
\td \right) \, \bar g_{ab}, \label{Rab}
\end{eqnarray}
where \be\lb{gage} F=\B-\A+(\td +1)\C+(d-1)\D. \ee

Using  the expressions for the Ricci tensor and substituting the
form field (\ref{SolF}), we  find three equations for $\B,\,\C$
and $\D$ with similar differential operators \bea
 \B'' +\B' F'  &=&
 \frac{\td b^2{\rm e}^{G-2F}}{2(D-2) }, \label{EqB}\\
 \C'' +\C' F'  &=&
-\frac{db^2{\rm e}^{G-2F}}{2(D-2) }+\sigma \td {\rm
e}^{2(\A-\C)},\lb{EqC}\\
\D'' +\D' F'  &=& \frac{\td b^2{\rm e}^{G-2F}}{2(D-2) },
\label{EqD} \eea where \be\lb{G} G=a\phi+2\B+2(d-1)\D,\ee and the
following  equation involving the function $A$: \bea (\A+
F)''-\A'(\A+ F)' + (\td+1)\C'^2 +&&\nonumber\\+(d-1) \D'^2 +
\frac12 \phi'^2 =\frac{\td b^2{\rm e}^{G }}{2(D-2) }.&& \eea
 The dilaton equation Eq.(\ref{dil}) takes the
following form
\begin{equation}\label{EqPhi}
\phi'' +\phi' F'  =
 \frac{a b^2{\rm e}^{G-2F}}2,
\end{equation}
where  primes denote derivatives with respect to $r$.

To simplify the above system we introduce a new independent
variable via \be\lb{taudef} d\tau=  \td \;\e^{-F} dr,\ee and pass
to a new function  \be \n=\A+F.\ee Then, denoting the derivatives
with respect to $\tau$  by dot, we obtain the following system:
\bea \ddot \B&=&\frac{b^2\,{\rm e}^{G}}{2\td (D-2)}\;,
\label{EqBt}\\\ddot D&=&\frac{b^2\,{\rm e}^{G}}{2\td (D-2)}\;,
\label{EqDt}\\\ddot \phi&=&\frac{ab^2\,{\rm e}^{G}}{2\td^2}\;,
\label{Eqfit}\\\ddot C&=&-\frac{b^2 d\,{\rm e}^{G}}{2\td^2
(D-2)}\;+\frac{\sigma}{\td}\; {\rm e}^{2(\n-\C)}, \label{EqCt}
\eea and \be \lb{EqAt}\ddot \n -{\dot \n}^2+{\dot \B}^2+(\td
+1){\dot \C}^2+(d-1){\dot \D}^2+\frac12{\dot
\phi}^2=\frac{b^2\,{\rm e}^{G}}{2\td (D-2)}\;.\ee Note that the
function $F$ has disappeared from the equations, showing $F$ to be
a gauge function. Under reparameterization of the radial
coordinate it can be chosen arbitrarily, in particular, set to
zero. Once the system (\ref{EqBt}-\ref{EqCt}) is solved,  the
function $\n$ can be expressed through $B\,,C\,,D$, and the Eq.
(\ref{EqAt}) becomes a constraint equation. Therefore the complete
solution for the metric functions and the dilaton should contain
eight free parameters subject to one constraint. One of them,
however, is redundant, since the the system is autonomous. In
return, we  have already introduced one free parameter $b$ related
to the form field strength (\ref{SolF}), so the actual number of
free parameters that we expect in the complete solution has to be
seven.

The Ricci scalar calculated for the solutions to the field
equations can be written in the following form \be\lb{Ricsc}
R=\frac12 {\phi'^2}\,\e^{-2A}
 +\frac{(d-\td)b^2}{2(D-2)}\, \e^{a\phi-2(\td +1)C} ,\ee
which is manifestly invariant under reparameterizations of the
radial coordinate.

\section{The complete solution}
The above system can be integrated as follows. First we observe
that the functions $\B,\,\D$ and $\td\phi/(a(D-2))$ may differ
only by a solution of the homogeneous equation, which is a linear
function of $\tau$, thus we obtain $D$ and $\phi$ in terms of $B$
as follows:
\bea\D&=&\B+d_1\tau+d_0,\lb{DB}\\\phi&=&\frac{a(D-2)}{\td}\B+
\phi_1\tau+\phi_0,\lb{FB}\eea where $d_0,\,d_1,\,\phi_0,\,\phi_1$
are free constant parameters. Substituting this into the Eq.
(\ref{G}) one finds  the following relation:\be \lb{GB}
G=\frac{\de (D-2)}{\td}B+g_1\tau+g_0,\ee where \be
\lb{Delt}\Delta=a^2+\frac{2d\td}{D-2},\ee and the integration
constants combine to \be g_{0,1}=a\phi_{0,1}+2(d-1)d_{0,1}. \ee
Together with the Eq. (\ref{EqBt}) one then obtains a decoupled
Liouville equation for $G$:\be\ddot G=\frac{b^2\de }{2\td^2}\;{\rm
e}^{G},\lb{LiuG}\ee from which the following first integral is
found straightforwardly: \be\lb{al} {\dot
G}^2-\frac{b^2\de}{\td^2}\;{\rm e}^{G}=\alpha^2,\ee with a new
integration constant $\alpha$  which can be real or pure imaginary
(for definiteness we will assume real $\alpha$ to be
non-negative). Note that under rescaling $\tau\to k^{-1}\tau$ the
parameter $\alpha$ will scale as $\alpha\to k\alpha$. The general
solution of (\ref{LiuG}) for real $\alpha\neq 0$ reads:
\be\lb{SolG} G=\ln\left(\frac{\alpha^2\td^2}{\de
b^2\,\sinh^2\left(\frac{\alpha}{2} (\tau-\tau_0)\right)}\right)
,\ee where $\tau_0$ is another integration constant. For
$\alpha=0$ one has instead \be\lb{SolG0}
G=\ln\left(\frac{4\td^2}{\Delta b^2(\tau-\tau_0)^2}\right).\ee For
imaginary $\alpha=i{\bar\alpha}$ the solution takes the form
\be\lb{SolGim} G=\ln\left(\frac{{\bar\alpha}^2\td^2}{\de
b^2\,\sin^2\left(\frac{{\bar\alpha}}{2}
(\tau-\tau_0)\right)}\right) ,\ee

Combining now the remaining equations, one can show that the
linear combination \be\lb{H} H=2(\n-\C)\ee obeys a second
decoupled Liouville equation \be\lb{EqH}\ddot H=2\s {\rm
e}^{H},\ee admitting the first integral \be \lb{Hint} {\dot H}^2
-4\s {\rm e}^{H}=\beta^2.\ee For $\sigma=1$ this parameter can be
real (in which case it will be assumed non-negative) or pure
imaginary, while for $\sigma=0,\,-1$ a solution exists only for
$\beta$ real. For real positive $\beta$ we find the following
solutions for all values of $\sigma$:
 \be \lb{SolH} H = \left\{
\begin{array}{ll}
 2\ln\beta/2-\ln\left[ \sinh^2(\beta \tau/2
   ) \right], \; & \sigma=1, \\
 \pm \beta \tau, & \sigma=0, \\
 2\ln\beta/2-\ln\left[ \cosh^2(\beta \tau/2
   ) \right], & \sigma=-1. \end{array}\right\}\ee
Note that we could introduce here another integration constant
replacing $\tau$ by $ \tau-\tau_1$ as in (\ref{SolG}), but since
the initial system is autonomous, one can choose without loss of
generality $\tau_1=0$ (while keeping $\tau_0$ in the solution for
$G$). For $\beta=0$ one has, for $\sigma=1$: \be\lb{SolH0} H=-\ln
\tau^2,\ee and for $\sigma = 0$ \be H = H_0 \ee constant, while
for pure imaginary $\beta=i{\bar\beta}$ and $\sigma=1$
\be\lb{SolHim}
H=\ln\left(\frac{{\bar\beta^2}}{4\sin^2\left({\bar\beta}\tau/2\right)}\right).
\ee Finally, expressing the metric functions $\A,\,\C$ from
(\ref{gage}), (\ref{H}), one can write the full solution in terms
of $G,\,H$ as follows: \bea
\B&=&\frac{\td}{\de(D-2)}\left(G-g_1\tau-g_0\right),\lb{Sol1}
\\
\D&=&\frac{\td}{\de(D-2)}\left(G-g_1\tau-g_0\right)+d_1\tau+d_0,\lb{Sol2}
\\ \C&=&\frac{1}{2\td}\;H-\frac{d}{\de
(D-2)}\; G+c_1\tau+c_0,\lb{Sol3}
\\ \A&=& \frac{(1+\td)}{2\td}\,H-\frac{d}{\de(D-2)}\;G-
F+c_1\tau+c_0,\lb{Sol5}\\
\phi&=&\frac{a}{\de}\;G+f_1\tau+f_0,\eea where\bea
c_{0,1}&=&\frac{a}{\de}\left(\frac{d}{D-2}\;\phi_{0,1}-
\frac{(d-1)a}{\td}\;d_{0,1}\right),
\nonumber\\
f_{0,1}&=&\phi_{0,1}-\frac{a }{\de}\;g_{0,1} =
   \frac{2\td}ac_{0,1}.\lb{Solc}\eea
The gauge function $F$ remains arbitrary and can be used to fix
the gauge in any convenient form. Our complete solution thus
depends on eight parameters: $b,
\,d_0,\,d_1,\,\phi_0,\,\phi_1,\,\tau_0,\, \alpha,\,\beta$, from
which the following four $ d_1,\, \phi_1, \, \alpha,\,\beta$ are
subject to  a constraint resulting from the Eq.
(\ref{EqAt}):\bea\lb{conspar}
&&\frac{(\td+1)\beta^2}{4\td}-\frac{\alpha^2}{2\de}-
 \td c_1^2  -(d-1)\left(\frac{dg_1}{\de(D-2)}-d_1\right)^2+ \nonumber\\
 &&+\left(\frac{\td  g_1}{\Delta (D-2)}\right)^2-
\frac{f_1^2}{2}=0.\eea Therefore, we have seven free parameters in
the generic solution. From this equation it follows that $\beta$
can be pure imaginary only if $\alpha$ is also imaginary, while
$\beta = 0$ is possible only for $\alpha$ imaginary or $\alpha =
0$ (the other parameters $d_1$ and $\phi_1$ also vanishing).

\section{Asymptotic flatness}
\subsection{Spherical transverse space}
Since the complete solution is determined by two functions
$G(\tau)$, $H(\tau)$ and a set of linear functions of $\tau$, one
has to investigate the behavior of the metric when $\tau\to 0$,
$\tau\to\tau_0$, and $\tau\to\pm\infty$. It is easy to see, that
for $\sigma=1$, in the limit $\tau\to 0$ we find a Minkowski space
after suitably choosing some free parameters.  The function $H$ in
this limit is \be\lb{Has}H\approx\ln\frac1{\tau^2},\ee while $G$
tends to a finite value \be G_0 =\ln\left(\frac{\alpha^2\td^2}{\de
b^2\sinh^2\left(\alpha\tau_0/2
 \right)}\right),\ee
if $\tau_0\neq 0$. Imposing the following two conditions on the
parameters
 \bea  d_0&=&0,\\a \phi_0 &=&G_0,\lb{phi0}\eea
 one obtains \bea B&\approx& 0,\\ D&\approx& 0,\\
 C&\approx& \ln\left(\frac1{|\tau|}\right)^{1/\td},\\
A&\approx& \ln\left(\frac1{|\tau|}\right)^{(1+1/\td)}.\eea
Choosing the gauge function to behave asymptotically as
\be\lb{Fas} F_\infty=\ln r^{\td+1},\ee one obtains from the Eq.
(\ref{taudef})\be\tau=-r^{-\td},\ee so transforming to the radial
variable $r$ we find the interval in the vicinity of the section
$\tau=0$ ($r \to \infty$) as \be ds^2 = - dt^2 + dx_1^2 + \cdots +
dx_p^2 + r^2 d\Sigma_{\td+1,1}^2 + dr^2.\ee The asymptotic value
of the dilaton is finite \be\phi_\infty=\phi_0.\ee The case
$\tau_0 = 0$, which leads to non-asymptotically flat solutions,
shall be examined elsewhere \cite{CGL}.

\subsection{Flat or hyperbolic transverse space: \\topological
censorship}

For $\sigma=0$ the quantity $H$ is a linear function of $\tau$, so
when $\tau\to 0$ all the metric functions go to constant values if
$\tau_0 \neq 0$.  Therefore the space-time is locally cylindrical,
\be ds^2 = - dt^2 + dx_1^2 + \cdots + dx_p^2 + r_0^2
d\Sigma_{\td+1,0}^2 + d\tau^2,\ee (after a suitable rescaling of
the world-volume coordinates) with $r_0$ constant, so that the
regular timelike section $\tau = 0$ is at finite distance. In the
hyperbolic case $\sigma=-1$ \be H\to \ln\frac{\beta^2}{4}\ee as
$\tau\to 0$, so the situation is similar. For $\tau \to \tau_0
\neq 0$, $G \to +\infty$, so that the section $\tau = \tau_0$ is
singular and again at finite distance.

Finally we inquire whether an asymptotic region, say $\tau =
+\infty$, can be a regular region at spatial infinity? When $\tau
\rightarrow +\infty$, $G \sim -\alpha\tau$. For asymptotic
flatness we require $B \to 0$ and $D \to 0$, leading to \be g_1 =
-\alpha, \quad d_1 = 0. \ee Then, \be c_1 = -\frac{\alpha
d}{\Delta(D-2)}, \ee so that \be \n \sim \frac{1+\td}{2\td}H \ee
goes to $+\infty$ only for $\sigma=0$ ($H \sim \beta\tau$).
Accordingly the metric asymptotes to \be ds^2 \sim -dt^2 + dx_1^2
+ \cdots + dx_p^2 + r^\frac{2}{1+\td}\,d\Sigma_{d+1,0}^2 + dr^2
\ee ($r \sim e^{\n} \sim e^{\frac{1+\td}{2\td}\beta\tau}$). This
is asymptotically flat only in the trivial case $\td =0$.

We conclude that the asymptotic flatness requirement cannot be
fulfilled for flat (toroidal) or hyperbolic transverse spaces, so
there are no asymptotically flat "topological" branes. In the
particular case of black holes $(d=1)$ this is a topological
censorship theorem which was proved recently \cite{CaGa01,Ga99}
for arbitrary space-time dimension with less assumptions than
here. Our argument shows that this is likely to be extendible to
the brane case as well.

\section{Horizons} From now on we will consider only spherical
transverse space. Horizons may arise in the asymptotic regions
$\tau\to\pm\infty$, when the metric function $\e^{2B}$ vanishes.
In these regions the function $G$ behaves (for $\alpha > 0$) as
\be \lb{Gas}G\approx-\alpha|\tau|+{\rm const},\ee so depending on
the ratio between parameters, one can get horizons in both
regions. Since the asymptotic region is at $\tau=0$ we will assume
that the outer horizon (the event horizon), if any, is located at
$\tau =-\infty$, and an internal Cauchy horizon (if any) --  at
$\tau=+\infty$.

In view of (\ref{Gas}), in the event horizon region one has
\be\lb{Bas}\e^{2B}\approx{\rm
const}\cdot\exp\left(\frac{2\td}{\Delta(D-2)}(\alpha-g_1)\tau\right).\ee
This goes to zero if \be \lb{neq1}\alpha>g_1.\ee Now we have to
distinguish the cases of non-degenerate and degenerate horizons.
Considering the prolongation of geodesics through the horizon, we
finds that in terms of the affine parameter $\lambda$, related to
$\tau$ via \be d\lambda=\e^{({\cal A}+B)}d\tau,\ee one has to
demand \be\lb{lamb}\e^{2B}\sim\lambda^n\ee on the horizon
($\lambda = 0$) where $n=1$ in the non-degenerate and $n\geq 2$ in
the degenerate case.

\subsection{Non-degenerate horizon}

In the case $n=1$ from the above reasoning we find the following
condition \be\e^{-(B+{\cal A})}\frac{d}{d\tau}\e^{2B}\to {\rm
const},\;\;{\rm as}\;\; \tau\to -\infty.\ee In view of the Eqs.
(\ref{Bas}), (\ref{neq1}) it is clear that $\dot B$ is non-zero
and
 \be\e^{\n}\sim\e^B,\ee
so that the exponential $\e^{{\cal A}}$ also vanishes on the
horizon. We then find that the Ricci scalar (\ref{Ricsc}) diverges
unless $\dot\phi$ vanishes in the limit $\tau\to -\infty$.
Demanding this quantity to  vanish we obtain the following
condition on the parameters: \be\lb{csol} c_1=-\frac{\alpha
a^2}{2\td\,\Delta}. \ee Now rewrite the constraint equation with
account for the definition of $H$ as follows \bea\lb{consn} && -
{\dot \n}^2+{\dot B}^2+(\td+1){\dot C}^2+(d-1){\dot D}^2
+\frac12{\dot \phi}^2\nonumber\\&& \qquad =
\frac{b^2}{2\td^2}\e^G-\frac{\td+1}{\td}\e^H.\eea It is easy to
see that both quantities on the right hand side of this equation
vanish on the horizon, while the first two terms on the left hand
side mutually cancel. From the positivity of the remaining part
one finds that $\dot C$ and $\dot D$ should also vanish on the
horizon. If we impose both these conditions then the constraint
equation on the parameters (\ref{conspar}) will be automatically
satisfied. Together with (\ref{csol}) this gives the following
relations \be\lb{condnondedg}
\alpha=\beta=-2d_1=-\frac{2\td}{a(D-2)}\phi_1. \ee Thus the
regularity of the event horizon fixes two more parameters (the
third follows from the constraint). It is then easy to check that
if $\tau_0>0$ our solution does not possess any singularity on the
semi-axis $-\infty<\tau<0$, so solutions are free from naked
singularities.

Of course, the behavior inside the horizon can be expected to be
singular, and one can show that the point $\tau=\tau_0$ is just
such a singularity. Substituting the solution obtained in the
vicinity of $\tau=\tau_0$ into the Eq. (\ref{Ricsc}) one obtains
\be \lb{sing0} R\sim \left(\frac1{\tau-\tau_0}\right)^{2\left(1+
\frac{2d}{\Delta (D-2)}\right)}. \ee It is convenient to choose
the Schwarzschild gauge which allows to parameterize the entire
space-time in a more transparent form. The corresponding gauge
function $F$ compatible with the asymptotic choice (\ref{Fas})
looks as follows \be\lb{Fall}
F=\ln\left(r^{\td+1}f_-f_+\right),\ee with \be f_\pm=1-
\frac{x_\pm}{x},\;\;x=r^\td,\;\;x_\pm=r_\pm^\td,\;\;x_-<x_+,\ee
where the positive values $r_\pm$ correspond to the location of
horizons in accord with previous assumptions. Indeed, integrating
the equation (\ref{taudef}) for $\tau$ we obtain
\be\tau=\frac1{x_+-x_-}\ln\frac{f_+}{f_-}.\ee By fixing the gauge
in the above way we have introduced two extra parameters $r_\pm$,
so now we can fix the scale, trading $\alpha$ for the difference
$x_+-x_-$: \be\lb{scale} \alpha=x_+-x_-.\ee In this case the
function $H$ will take the simple form for $x>x_+$ and $x<x_-$:
\be H=\ln(x-x_+)(x-x_-).\ee Note that the function $\tau(x)$
becomes complex in the region $x_-<x<x_+$ between the horizons,
but this does not make the solution in terms of $x$ complex.
Passing through the horizon the variable $\tau$ shifts into the
complex plane by $i\pi/\alpha$, which is precisely what is needed
to account for the necessary sign change of the exponential
$\e^G$, positive for real $\tau$, and negative for $\tau = {\rm
  Re}(\tau) + i\pi/\alpha$. Thus the outer region $r>r_+$
maps to the half of the real axis $-\infty<\tau<0$, the interval
$r_-<r<r_+$ maps to the line in the complex plane of $\tau$
parallel to the real axis and shifted by $i\pi/\alpha$, with
 the point $r_-$ corresponding to
${\rm Re}(\tau)=\infty$,  while the region $r<r_-$ (if the section
$r = r_-$ is non-singular, which is possible only in the absence
of the dilaton) maps to the part of the positive real axis of
$\tau$ from infinity to $\tau= \tau_0$. Let $x_0=r_0^\td$ be the
image of $\tau_0$ (we assume that $r_0<r_-$). In terms of these
quantities the function $G$ can be presented as follows \be
\e^G=\frac{4\td^2(x_+-x_0)(x_--x_0)}{\Delta b^2}f_+f_-f_0^{-2},\ee
where \be f_0=1-\frac{x_0}{x}.\ee With this parametrization the
solution now takes the following form \bea\lb{solfx0}
ds^2&=&\left(\frac{f_-}{f_0}\right)^{\frac{4\td}{\Delta(D-2)}}
\left(-\frac{f_+}{f_-}dt^2+d{\bf x}^2\right)
+\nonumber\\&+&f_-^\frac{2a^2}{\Delta\td}f_0^\frac{4d}{\Delta(D-2)}
\left(r^2d\Sigma_{k,1}+\frac{dr^2}{f_+f_-}\right),\\
\e^{\phi-\phi_0}&=&\left(\frac{f_-}{f_0}\right)^\frac{2a}{\Delta}.\eea
Here $\phi_0$ is given by Eq. (\ref{phi0}) as a function of other
parameters $b,\,\tau_0$, so totally the solution seems to depend
on four free parameters $b,\,r_\pm,\,r_0$. This differs from the
standard black brane solution (with an arbitrary value of the
dilaton at infinity, usually assumed zero) only by the presence of
the factors $f_0$. According to the Eq. (\ref{sing0}) the point
$r=r_0$ is a curvature singularity, while, as it easy to check,
the point $r=0$ is not. So with our choice of coordinates the
central singularity is at $r=r_0$, and by the coordinate
transformation $x\to x+x_0$ or \be r\to
\left(r^\td+r_0^\td\right)^{1/\td},\ee it can be moved to $r=0$.
This amount to set $x_0=0$ in (\ref{solfx0}), and we arrive at the
standard form of the black brane solution \cite{DuLu94}
\bea\lb{solf} ds^2&=&
{f_-}^{\frac{4\td}{\Delta(D-2)}}\left(-\frac{f_+}{f_-}dt^2+d{\bf
x}^2\right)+\nonumber\\&+&f_-^\frac{2a^2}{\Delta
\td} \left(r^2d\Sigma_{k,1}+\frac{dr^2}{f_+f_-}\right),\\
\e^{\phi-\phi_0}&=& {f_-}^\frac{2a}{\Delta},\eea with only three
free parameters.

\subsection{Degenerate horizon}
For a double horizon ($n = 2$ in (\ref{lamb})), we have near the
horizon \be \e^{2B} \sim \lambda^2\,, \quad
\frac{d\e^{2B}}{d\lambda} = \e^{B-\n}\dot{B} \sim \lambda\,, \ee
leading to \be\lb{nb1} \e^{-\n}\dot{B} \sim 1. \ee A first
possibility is that both $\dot{B}$ and $\e^{\n}$ are finite on the
horizon, implying $\dot{\n} \sim 0$. However the constraint
equation (\ref{consn}) then implies, in particular, that $\dot{B}
= 0$, contrary to the assumption. So the solution of (\ref{nb1})
is that both $\dot{B}$ and $\e^{\n}$ vanish on the horizon. The
first condition implies \be\alpha=g_1,\ee then the condition
$\e^B=0$ can only be satisfied if $\alpha=0$, in which case the
solution for $G$ is given by (\ref{SolG0}). We then see from the
second condition $\e^{\n}\sim0$ that the Ricci scalar
(\ref{Ricsc}) diverges on the horizon unless $\dot \phi \sim 0$.
This condition fixes the value of the constant $ f_1=0$. Together
with the previous condition $g_1=0$ this gives \be
d_1=\phi_1=0.\ee Now from the constraint equation (\ref{conspar})
we see that $ \beta=0$ so the solution for $H$ must be taken in
the form (\ref{SolH0}) (and $\sigma=1$). Finally, we obtain the
conditions \be\alpha=\beta=d_1=\phi_1=0,\ee which, together with
our choice of a scale (\ref{scale}) naturally fit the previous
conditions (\ref{condnondedg}) in the limit \be r_-\to r_+.\ee

\section{Cosmological constant}
Here we investigate the same system in presence of the
cosmological constant. For simplicity we set the dilaton to zero
since already the black hole case leaves a little hope to find
analytic solutions in this case. The system of equations takes the
following form (in $r$-terms):\bw
 \bea
 \B'' +\B' F'  &=&
 \frac{\td b^2}{2(D-2)}\;{\rm e}^{2\left(\A-(\td+1)\C\right)}+
 \Lambda {\rm e}^{2\A}, \label{EqBl}\\
 \C'' +\C'F' &=&
-\frac{d b^2}{2(D-2)}\;{\rm e}^{2\left(\A-(\td+1)\C\right)} +
\sigma \td {\rm e}^{2(\A-\C)}+\Lambda {\rm e}^{2\A},\lb{EqCl}\\
\D'' +\D'F' &=& \frac{d b^2}{2(D-2)}\;{\rm
e}^{2\left(\A-(\td+1)\C\right)}
+\Lambda {\rm e}^{2\A}, \label{EqDl} \\
B''+B'(B'-A')&+&(\td+1)\left[C''+C'(C'-A')\right]+
(d-1)\left[D''+D'(D'-A')\right]=\nonumber\\&=&\frac{\td
b^2}{2(D-2)}\;{\rm e}^{2\left(\A-(\td+1)\C\right)}+\Lambda {\rm
e}^{2\A}, \lb{EqAl}\eea \ew where no specific gauge was imposed.

 Note that our ansatz for
$\sigma=0,\,-1$ corresponds to 'topological' solutions, so our aim
was to explore whether topological black holes \cite{Le00} admit
$p$-brane generalizations. The main difference in the structure of
the systems (\ref{EqBl}--\ref{EqAl}) and  the system considered in
Sec. 2 is that now one has three different exponentials instead of
two for zero cosmological constant. So, with some luck, one could
hope  to obtain three Liouville equations. We could proceed along
the same lines choosing e.g. the gauge $F = 0$, but this does not
lead to decoupling in the general case: we were not able to reduce
the resulting system to a set of Liouville equations or to
integrate it in any other way except for the case $d=1$
corresponding to black holes. So to answer the question about
existence of topological branes numerical work is likely to be
required. This is beyond the scope of the present paper, so we
will proceed in obtaining a complete solution for the case $d=1$
only. We obtain the solution which was known earlier, but contrary
to previous treatment we find it via a complete integration of the
corresponding system of equations, thus providing the uniqueness
proof for topological zero-branes with homogeneous transverse
space.

For $d=1$ the $\D$-part of the equations disappears, and assuming
the curvature gauge \be\C=\ln r,\ee we are left with the system\bw
\bea
 \B'' +\B' (\B-\A')+\frac{\td+1}{r}\;\B'  &=&
 \left(\frac{\td b^2}{2(D-2)r^{2(\td+1)}}+
 \Lambda\right) {\rm e}^{2\A}, \label{EqBl1}\\
 \frac{\td}{r^2}+\frac1{r}\;(B'-A')&=&\left(\sigma \td-
 \frac{ b^2}{2(D-2)r^{2\td}}+
 \Lambda\right) {\rm e}^{2\A},\lb{EqCl1}\\
B''+B'(B'-A')-\frac{\td+1}{r}\;\A' &=& \left(\frac{\td
b^2}{2(D-2)r^{2(\td+1)}}+
 \Lambda\right) {\rm e}^{2\A},\lb{EqAl1}\eea \ew
Comparing the Eqs. (\ref{EqBl1}) and (\ref{EqAl1}) we see that \be
\B=-\A+ {\rm const}, \ee where the integration constant can be
removed by rescaling of time. Then the Eq. (\ref{EqCl1}) becomes a
first order decoupled equation for $\A$ which can be rewritten
as\be r^{1-\td} \frac{d}{dr} \left(r^\td {\rm
e}^{-2\A}\right)=\sigma\td-
\frac{b^2}{2(D-2)r^{2\td}}+r^2\Lambda\;.\ee The solution reads
\be{\rm
e}^{-2\A}=\sigma-\frac{2M}{r^\td}+\frac{Q^2}{r^{2\td}}+\frac{\Lambda
r^2}{\td+2},\ee where \be Q^2=\frac{b^2}{2\td(D-2)},\ee and $M$ is
the integration constant. For $\Lambda<0$ and $\sigma=0, -1$ this
is the standard two-parametric family of charged topological black
holes in $D$ dimensions. Thus, by  complete integration of the
field equations we prove that this solution is unique within the
class considered.

\section{Conclusions}
In this paper, we have constructed the general solution to the
metric-dilaton-antisymmetric form action in arbitrary dimensions
describing black $p$-branes with homogeneous transverse spaces of
different topologies. The solution contains  eight free parameters
subject to a constraint involving four of them. When the
asymptotic flatness condition is imposed on solutions with
spherical transverse space, the number of independent parameters
is reduced by two. The requirement of the existence of a regular
horizon further reduces this number by two or three depending on
whether the horizon is non-degenerate or not. Finally we are left
with the standard solutions which are determined by their mass and
charge densities and do not contain any extra parameters (apart
from the asymptotic value of the dilaton). Our result proves the
uniqueness of the standard asymptotically flat $p$-brane solutions
with spherical transverse space for both non-degenerate and
degenerate event horizons. (In particular, we extend the
uniqueness proof for the multidimensional static dilatonic black
holes \cite{Gi02} in the case of a degenerate horizon, assuming
spherical symmetry). The only physical parameters of a singly
charged $p$-brane are its mass and charge densities and the
asymptotic value of the dilaton.
 All extra parameters arising in the course
of complete integration of the field equations are removed by
requiring asymptotic flatness and absence of naked
singularities. The $p$-brane solutions with extra parameters
previously suggested in the literature (the three-parameter
solution of \cite{LuPoXu95}, the four-parameter solution of
\cite{ZhZh99}, the solutions discussed in
\cite{BeVe00,BeLe00,Lo01,Ba01}) therefore  are not free of naked
singularities, unless the extra parameters are given some
particular values.

We have also obtained the complete solutions for $p$-branes with
hyperbolic or toroidal transverse spaces, but these were
found to be incompatible with the requirement of asymptotic
flatness. In fact, this is what could be expected, since in the
better studied case of black holes the topological censorship
theorem forbids asymptotically flat solutions with non-trivial
topology of the horizon. Our considerations prove topological
censorship for $p$-branes, assuming homogeneity of the transverse
space, and we believe that this conjecture is true under less
restrictive assumptions as well.

Finally we attempted to find general black brane solutions in the
presence of a cosmological constant, but the system of equations
is not fully integrable in this case. We were able to find the
complete solution only in the case $d=1$, i.e. for the black hole.
For a negative cosmological constant the solutions exist for all
three topologies of the event horizon and coincide with the
multidimensional topological black holes previously presented in
\cite{CaGa01}. Thus our complete integration procedure proves the
uniqueness of charged topological black holes in arbitrary
dimensions under the assumption of homogeneity of the transverse
space.

\begin{acknowledgments}
D.G. is grateful to CENTRA/Instituto Superior T\'ecnico (Lisbon)
and GTAE for hospitality and NATO for support in August 2002 when
this paper was initiated.  He also thanks LAPTH Annecy for
hospitality and support in December 2003 when the paper received
its final form. JPSL acknowledges a grant from FCT through the
project ESO/PRO/1250/98.

\end{acknowledgments}

\end{document}